\documentclass[aps,pra,nofootinbib,floatfix,preprintnumbers,tightenlines,11pt,superscriptaddress]{revtex4-2}

\usepackage{amssymb}
\usepackage{amsthm}
\usepackage{amsmath}
\usepackage{verbatim}
\usepackage{algorithmic}
\RequirePackage[OT1]{fontenc}
\RequirePackage{hyperref}

\usepackage{fullpage}
\usepackage{amsmath,amssymb}
\usepackage{theoremref}
\usepackage{graphicx}
\usepackage{bm,mathrsfs,amsfonts}
\usepackage{algorithm2e}
\usepackage{bbm}
\usepackage[section]{placeins}

\usepackage{hyperref}
\hypersetup{
     colorlinks   = true,
     citecolor    = blue
}

\usepackage{float}
\graphicspath{ {./figures/} }
\usepackage[dvipsnames]{xcolor}
\usepackage{titlesec}
\usepackage{comment}

\input{Qcircuit}

\DeclareMathOperator{\im}{im}						% Image
\newcommand{\R}{\mathbb{R}}							% Real Numbers
\newcommand{\Z}{\mathbb{Z}}							% Integers.
\renewcommand{\dim}{\mathbbm{d}}					% Dimension.
\newcommand{\K}{\mathcal{K}}						% A simplicial complex
\newcommand{\goto}{\rightarrow}						% Function mapping ->
\newcommand{\del}[1]{\partial_{#1}}					% Boundary maps
\newcommand{\LC}{\left\{} \newcommand{\RC}{\right\}}% Curly Braces { }
\usepackage{subfigure}
\newcommand{\norm}[1]{\left\lVert #1 \right\rVert}
\newtheorem{defn}{Definition}%[section]

\usepackage[scr=boondoxo]{mathalfa}

\newcommand{\be}{\begin{equation}}
\newcommand{\ee}{\end{equation}}
\newcommand{\ba}{\begin{align}}
\newcommand{\ea}{\end{align}}
\newcommand{\ra}{\rangle}

\newcommand{\eps}{\epsilon}

\newcommand{\HH}{\mathcal{H}}

\renewcommand{\a}{\alpha}

\newcommand{\s}{\sigma}

\renewcommand{\l}{\lambda}

\begin{document}
\title{Quantum Persistent Homology}

%\author{Bernando Ameneyro  \thanks{Department of Mathematics, University of Tennessee Knoxville}\footnotemark[3]
%\and George Siopsis  \thanks{Department of Physics, University of Tennessee Knoxville}
%\and Vasileios Maroulas \footnotemark[3] 
%}

%\maketitle

\author{Bernardo Ameneyro}
\email{bameneyr@vols.utk.edu}
\affiliation{Department of Mathematics, The University of Tennessee, Knoxville, TN 37996-1320, USA}
\author{Vasileios Maroulas}
\email{vasileios.maroulas@utk.edu}
\affiliation{Department of Mathematics, The University of Tennessee, Knoxville, TN 37996-1320, USA}
\author{George Siopsis}
\email{siopsis@tennessee.edu}
\affiliation{Department of Physics and Astronomy, The University of Tennessee, Knoxville, TN 37996-1200, USA}
\date{\today}

\begin{abstract}
Persistent homology is a powerful mathematical tool that summarizes useful information about the shape of data allowing one to detect persistent topological features  while one adjusts the resolution. However, the computation of such topological features is often a rather formidable task necessitating the sub-sampling the underlying data. To remedy this, we  
develop an efficient quantum computation of persistent Betti numbers, which track topological features of data across different scales. Our approach employs a persistent Dirac operator whose square yields the persistent combinatorial Laplacian, and in turn the underlying persistent Betti numbers which capture the persistent features of data. We also test our algorithm on point cloud data.  
\end{abstract}

\maketitle

\section{Introduction}

%{\color{red}We can do much better than this. Look at my paper. I have 40 citations in the Introduction. Do something similar.}

% DA intro
%These days mankind produces insane amounts of data, which can yield very useful information if analyzed correctly. For instance, in the medical field, many brain-computer interfaces require accurate methods for the classification of electroencephalogram (EEG) signals \cite{garrett2003comparison}. A topological approach was used to analyze breast cancer data and identify a subgroup with a special mutational profile and excellent survival chances \cite{nicolau2011topology}. 

% TDA Intro and how it works
%Topological data analysis (TDA) is a powerful technique for extracting information on topological features of data sets and analyzing how they change over different scales {\color{red} Cite papers}. Topological information is particularly useful because it is independent of the representation of the data and the presence of noise. Persistent homology is a TDA tool that studies nested collections of simplicial complexes (filtrations) and how topological features such as connected components, holes, and voids appear and disappear as the scale changes \cite{zomorodian2005computing}.

In recent years, topology and geometry blended with statistical methods have seen increasing application to the study of data analysis, visualization, and dimensionality reduction \cite{carlsson,carlsson_basics,CompyTopo,vasudevan2013,kusano2016,edelsbrunner2013,chung2015,munch2013,munch2016}. These applications range from classification and clustering in fields such as action recognition \cite{tda_action}, handwriting analysis \cite{tda_number}, and biology \cite{tda_wheeze,tda_clustering2015,gunnarcancer, HodgeCycle,CPD_me}, to classification of high entropy alloy  \cite{Maroulas2020} and gas separation \cite{Townsend2020}, to the analysis of complex biological networks \cite{Maroulas2021}, and other complex dynamical systems \cite{tda_windows,tda_tracking} and sensor networks \cite{GdS06,GdS07,Ghrist12,CdS1,CdS2}. 

Persistent homology \cite{persissensor, jholes, fullerene,tda_signal,persis_brain}, the workhorse of topological data analysis (TDA), has helped to compress nonlinear point cloud data from a new geometrically faithful point of view. In the realm of signal analysis, classification and clustering based on topological features of the signal identifies features that traditional signal analysis techniques fail to detect \cite{MaMa16,Marchese2018,tda_action,tda_timeseries,tda_wheeze}. Topological features, such as the number of connected components, cycles, and higher-dimensional voids, respectively, represent multi-stability, periodicity, and chaos in a dynamical system \cite{tda_wheeze, tda_windowgenes}.

The effective computation of persistence diagrams has become an area of intense active research, including a significant successful effort toward facilitating previously challenging computations. For example, the creation of persistence diagrams with packages such as Dionysus \cite{dionysus} and Ripser \cite{ripser} take advantage of certain properties of simplicial complexes \cite{persistwist}. Point clouds typically consist of many points. A set of $n$ such points possesses $2^n$ potential subsets that could contribute to the topology. The best classical algorithm for estimating these homological features of a data set with accuracy $\alpha$ takes time $O(2^n \log (1/\alpha)).$ 

More recently, quantum algorithms that compute topological features of data were developed. The first such algorithm was introduced in \cite{lloyd2016quantum} with run time $O(n^5/\alpha)$ -- exponentially faster than the best known classical algorithms. This algorithm was designed for a discrete-variable (DV) quantum system based on qubits, and it was extended to a continuous-variable (CV) substrate in \cite{siopsis2019quantum}. These two algorithms compute Betti numbers by analyzing a linear operator called the Dirac operator whose square is the combinatorial Laplacian. However, as was noted in \cite{gunn2019review}, these algorithms do not compute \emph{persistent} topological features so that one may track how topological features persist as the resolution of data changes and the underlying noise may vary. Recently, a quantum computation of persistent Betti numbers based on block-encoding and quantum singular value transformation was introduced in \cite{hayakawa2021quantum}. However, the study did not address any real implementation on data. 
To that end, a few recent attempts on data for tracking the non-persistent topological features were studied. Indeed, the work in \cite{Huang2018}  demonstrates a quantum algorithm presented in \cite{lloyd2016quantum} by employing a six-photon quantum processor to successfully analyze the topological features of a network including three data points. Moreover the work in \cite {1Qb} developed  a quantum annealing approach for topologically analyzing point cloud data, and \cite{wie2017quantum} gives a quantum circuit to construct all maximal cliques of an $n$-node network using Grover's search algorithm. 
%Other quantum algorithms for TDA were also developed \cite{1Qb, wie2017quantum, Huang2018}. {\color{blue}Add one line details on [42,43] } 

The persistent combinatorial Laplacian is expressed as a quadratic function of a pertinent linear map called the boundary operator. However, it is computationally efficient to consider a map linear on the boundary. Here, we introduce a quantum algorithm that relies on a (persistent) Dirac operator which is linear in the boundary operator. Its square yields the persistent combinatorial Laplacian, hence the persistent Betti numbers, allowing us to track topological features across different scales and obtain persistence information. To that end, our algorithm is a generalization of the algorithm introduced in \cite{lloyd2016quantum}, and it similarly  offers an exponential speedup over known classical algorithms of persistent homology.

Our discussion is organized as follows. In Section \ref{sec:2}, we review persistent homology  introducing the important concepts relevant to the algorithm, such as the persistent combinatorial Laplacian and persistent Betti numbers. In Section \ref{sec:3}, we present our quantum algorithm introducing the persistent Dirac operator together with an outline of the quantum tools that are needed, such as quantum random access memory (QRAM), Grover's search algorithm, and phase estimation. In Section \ref{sec:4}, we apply our algorithm to a data set whose points are organized in squares, which was suggested in \cite{gunn2019review} as a case in which persistent Betti numbers differ from Betti numbers. Finally in Section \ref{sec:5}, we offer concluding remarks. Details of subroutines needed for our quantum algorithm are provided in Appendices \ref{app:A}, \ref{app:B}, and \ref{app:C} (Grover's algorithm, implementation of an exponential operator, and phase estimation, respectively).

% Classical issues
%Unfortunately,  classical algorithms of TDA require resources and computational time that scale exponentially with the size of data. Execution of these algorithms presents serious challenges even to the most powerful high performance classical computers. Quantum algorithms present an alternative promising significant, often exponential, speedup, and have attracted considerable attention {\color{red} Cite papers}. In the context of TDA, Lloyd, Garnerone, and Zanardi showed that it is possible to attain exponential speedup under certain circumstances using a qubit architecture \cite{lloyd2016quantum}. Their method was shown to be applicable to substrates based on continuous variables (qumodes) that can be realized with today's available technology running at room temperature \cite{siopsis2019quantum}.

% Previous work
%We follow the work of \cite{lloyd2016quantum} which present quantum algorithms that compute the Betti numbers of a data set. That is, they extract topological information from the data at a fixed scale. 

% Differences with new paper

%\section{Preliminary Background Information}

\section{Persistent Homology}\label{sec:2}

% Big picture explanation of the algorithm
Persistent homology studies objects called simplicial complexes. The classical algorithms that perform topological data analysis use the data to construct simplicial complexes, e.g., by connecting all the points in a point cloud within a certain distance from each other (Vietoris-Rips complex) and then varying the distance to obtain a filtration of simplicial complexes. After that, the algorithms proceed to compute the eigenvalues and eigenvectors of linear operators, such as the boundary operator and the persistent combinatorial Laplacian that act on the complexes.

Next, we briefly discuss simplicial complexes and homology, an algebraic descriptor for coarse shape in topological spaces. In turn, persistent homology, and its summary, persistence diagrams, are techniques for bringing the power and convenience of homology to the description of subspace filtrations of topological spaces.
We consider topological spaces of discernible dimension called manifolds.

\begin{defn}
A topological space $X$ is called a $k-$dimensional manifold if every point $x \in X$ has a neighborhood which is homeomorphic to an open neighborhood in $k$-dimensional Euclidean space.
\end{defn}

We generalize the fixed-dimension notion of a manifold in order to define simplicial homology for simplicial complexes.
We then discuss the Vietoris-Rips construction which is used to associate simplicial complexes to datasets. 

\begin{defn} \label{simplex}
A $k$-simplex is a collection of $k+1$ linearly independent vertices along with all convex combinations of these vertices:
\begin{equation} \label{convex_combo}
\s=[ v_0,...,v_k ] = \LC \sum_{i=0}^k \alpha_i v_i : \sum_{i=0}^k \alpha_i = 1 \textrm{ and } \alpha_i \geq 0 \, \forall i \RC.
\end{equation}
Topologically, a $k$-simplex is treated as a $k$-dimensional manifold (with boundary).
An oriented simplex is typically described by a list of its vertices, such as $[ v_0, v_1, v_2 ]$.
The faces of a simplex consist of all the simplices built from a subset of its vertex set;
for example, the edge $[v_1, v_2]$ and vertex $[v_2]$ are both faces of the triangle $[ v_0, v_1, v_2 ]$. 
\end{defn}

\begin{defn} \label{simplicial_complex}
A simplicial complex $\K$ is a collection of simplices wherein \\
(i) if $\sigma \in \K$, then all its faces are also in $\K$, and \\
(ii) the intersection of any pair of simplices in $\K$ is another simplex in $\K$. \\
We denote the collection of $k$-simplices within $\K$ by $\K_{k}$. 
\end{defn}
Conditions (i) and (ii) in Definition \ref{simplicial_complex} establish a unique topology on the realization of a simplicial complex which restricts to the subspace topology on each open simplex.
For finite simplicial complexes realized in $\R^{\dim}$, this topology is also consistent with the Euclidean subspace topology.

Here we define the homology groups for a simplicial complex through purely combinatorial means, which allows for automated computation. 
\begin{defn} \label{defn_chain_group}
The chain group (over $\Z_3$) on a simplicial complex $\K$ of dimension $k$ is denoted by $C_k(\K)$ and is defined as formal sums of $k$-simplices in $\K$:
\begin{equation} \label{eqn_chain_group}
C_k(\K) = \LC \sum_{\sigma \in \K_{k}} n_\sigma \sigma : n_\sigma \in \Z_{3} \RC.
\end{equation}
\end{defn}

\begin{defn} \label{defn_boundary_map}
The $k$-th boundary map is a homomorphism $\del{k} : C_k(\K) \goto C_{k-1}(\K)$ defined on each simplex as an alternating sum over the faces of dimension $k-1$:
\begin{equation} \label{eqn_boundary_map}
\del{k}[v_0,...,v_k] = \sum_{n=0}^k (-1)^n [v_0,...,v_{n-1},v_{n+1},...,v_k].
\end{equation}
\end{defn}

%\begin{remark}
Chain groups give an algebraic way to describe subsets of simplices as a formal sum.
Toward this viewpoint, the chain group is often defined over $\Z_2 = \LC 0 , 1 \RC$ instead of $\Z_{3}$.
In this case, the boundary maps can be understood classically; e.g., the boundary of a triangle yields (the sum of) its three edges and the boundary of an edge yields (the sum of) its endpoints. 
When viewed over $\Z_{3}$, the presence of sign specifies simplex orientation. 
%\end{remark}

Putting chain groups of every dimension together along with the boundary maps successively defined between them, we obtain a chain complex:
\begin{equation} \label{eqn_chain_complex}
\LC 0 \RC \xleftarrow{\bm 0} C_0(\K) \xleftarrow{\del{1}} C_1(\K) \xleftarrow{\del{2}} C_2(\K) \xleftarrow{\del{3}} ... 
\end{equation}
The composition of subsequent boundary maps yields the trivial map \cite{CompyTopo};
this property is typically rephrased as $\im(\del{k+1}) \subset \ker(\del{k})$ which enables definition of the following modular groups.
\begin{defn} \label{defn_homology_group}
The homology group of dimension $k$ is given by
\begin{equation} \label{eqn_homology_group}
H_k(\K) = \ker(\del{k}) / \im(\del{k+1}) = \LC \tilde{x} = x + \im(\del{k+1}) : x \in \ker(\del{k}) \RC\!,
\end{equation}
where $\tilde{x} = \LC x + y : y \in \im(\del{k+1}) \RC$ defines the coset equivalence class of $x$.
\end{defn}
The dimension of $H_{k}$ is the $k$-th Betti number, denoted by $\beta_{k}$. It is shown in \cite{wang2020persistent} that the $k$-th Betti number is also the dimension of the kernel of the $k$-th combinatorial Laplacian $\mathcal{L}_{k}^{\K}:C_{k}(\K)\to C_{k}(\K)$ given by $\mathcal{L}_{k}^{\K} = \partial_{k}^{\dagger}\partial_{k} + \partial_{k+1}\partial_{k+1}^{\dagger},$
where $\partial_{k}^{\dagger}$ denotes the dual.

The generators of the homology group correspond to topological features of the complex $\K$;
for example, generators for the $0$-homology group correspond to connected components, generators of $1$-homology group correspond to holes in $\K$, etc. 
The interpretation of these features is exemplified by taking the topological boundary of a $k+1$ ball (that is, a $k$-sphere);
for example, the boundary of an interval is two (disconnected) points while the boundary of a disc is a loop. 

We wish to extend the notion of homology for a discrete set of data $\bm x = \LC v_i \RC_{i=1}^N$ within a metric space $(X,d_X)$.
Treating the set itself as a simplicial complex, its homology yields only the cardinality of the data points.
So, we utilize the metric to obtain more information.
Here we denote by $B(x_0,r_0)$ a metric ball centered at $x_0$ of radius $r_0$. 
Fix a radius $r > 0$ and consider the collection of neighborhoods $U = \LC U_i \RC = \LC B(v_i,r) \RC$ along with its union $\mathcal{U}_r = \cup_i B(v_i,r)$.
The collection of sets $\LC \mathcal{U}_r \RC_{r \in \R^+}$ naturally yields information about the arrangement within $X$ of the dataset $\bm x$ at various scales. 
To make homology computations more tractable for $\mathcal{U}_r$, people often consider instead the associated Vietoris-Rips (VR) complexes which is (topologically) very similar. %{\color{red}Do we need to cite something here?}.
%\begin{defn} \label{defn_nerve_and_cech}
%The nerve $\mathcal{N}(U)$ of a collection of open sets $U$ is the simplicial complex where a $k$-simplex $[v_{i_0},..., v_{i_k}]$ is in $\mathcal{N}(U)$ if and only if $\cap_{j=0}^k U_{i_j} \neq \emptyset$. 
%The nerve of the neighborhoods $U = \LC B(v_i,r) \RC$ is called the $\cech$ complex on the data $\LC v_i \RC$ at radius $r$ and is denoted by $\textrm{\v Cech}(\bm x, r)$. 
%\end{defn}
\begin{defn} \label{defn_vr}
The Vietoris-Rips complex of the data $\LC v_{i}\RC$ at scale $\eps$, denoted by $S^{\eps}(\bm x)$ (or simply $S^\eps$), is the simplicial complex where a $k$-simplex $\s=[v_{i_0},..., v_{i_k}]$ is in $S^{\eps}(\bm x)$ if and only if $\text{diam}(\s) = \max_{j,j'}\LC d_{X}(v_{i_j},v_{i_{j'}})\RC\le \epsilon$.
\end{defn}

\begin{figure}
\begin{center}
\includegraphics[width=1.6in,height=1.4in]{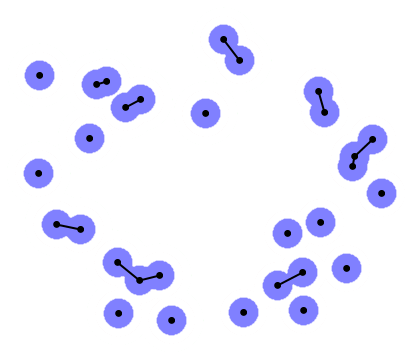} 
\includegraphics[width=1.6in,height=1.4in]{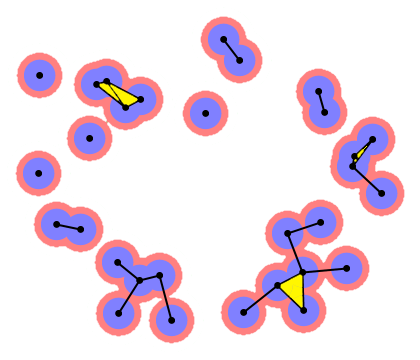} 
\includegraphics[width=1.6in,height=1.4in]{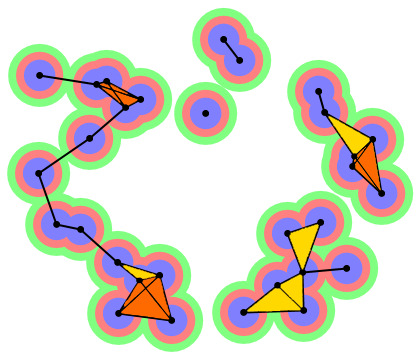}
\end{center}
\caption{The neighborhood space and VR complex of matching scale plotted at three different scales. Yellow indicates a triangle while orange indicates a tetrahedron. This family of simplicial complexes is the filtration utilized to compute and define persistent homology.} \label{grow_cech}
\end{figure}

Examples of the VR complex for the same data at different scales are depicted in Fig. \ref{grow_cech}, where they are superimposed with the associated neighborhood space.
% {\color{red}Any nerve complex trivially satisfies the requirements for a simplicial complex \cite{CompyTopo}.
% Moreover, the nerve theorem states that the nerve and union of a collection of convex sets have similar topology (they are homotopy equivalent) \cite{Hatcher};
% specifically, the $\cech$ complex and neighborhood space $\mathcal{U}$ have identical homology for any given radius.} 

Unfortunately the tools defined so far would only allow to obtain the Betti numbers of a fixed scale $\eps$, which is not very useful to analyze data because there is no way to differentiate between true topological features and noise. To tell them apart one must track the topological features across scales, those that persist for a long time are the true topological features, while those that persist for a short time are just noise. So, we must extend the notions above to include multiple simplicial complexes at once.

Consider a nested sequence or filtration of simplicial complexes
\begin{equation} \label{eqn_filtration}
\emptyset \subseteq \K_{1} \subseteq \K_{2} \subseteq ... \subseteq \K_{n} 
\end{equation}
each complex $\K_{i}$ has associated chain groups $C_{k}(\K_{i})$, boundary operators $\partial_{k}^{i}$ and Homology groups $H_{k}^{i}$ as before. But now the inclusion maps $\iota:\K_{i}\goto\K_{j}$ between complexes induce homomorphisms $h_{k}^{i,j}=\iota_{*}:C_{k}(\K_{i})\goto C_{k}(\K_{j})$ between their corresponding chain groups.
\begin{defn} \label{defn_persist_hom_group}
For two nested simplicial complexes $\K_{i}\subseteq\K_{j}$ we define their $k$-th persistent Homology group $H_{k}^{i,j}$ as
\begin{equation}
H_{k}^{i,j} = \text{Im} (h_{k}^{i,j}) =  \text{Ker}(\partial_{k}^{i})/(\text{Im}(\partial_{k+1}^{j}) \cap \text{Ker}(\partial_{k}^{i}))
\end{equation}
where $\text{Ker}(\partial_{k}^{i})$ is viewed as a subgroup of $\text{Ker}(\partial_{k}^{j})$. The dimension of $H_{k}^{i,j}$ is the $k$-th persistent Betti number $\beta_{k}^{i,j}$.
\end{defn}

Let $\tilde{C}_{k+1}^{i,j} = \LC x\in C_{k+1}(\K_{j}) : \partial_{k+1}^{j} x \in C_{k}(\K_{i})\RC$, that is, the subgroup of $C_{k+1}(\K_{j})$ defined by the $k+1$-simplices in $\K_{j}$ with boundary in $\K_{i}$. Then we can define the $k$-th persistent combinatorial Laplacian
\begin{equation}\label{eq:6}
\mathcal{L}_{k}^{i,j} = \tilde{\partial}_{k}^{i,i\dagger} \tilde{\partial}_{k}^{i,i} + \tilde{\partial}_{k+1}^{i,j} \tilde{\partial}_{k+1}^{i,j\dagger},
\end{equation}
where $\tilde{\partial}_{k+1}^{i,j}$ is the restriction of the boundary operator $\partial_{k+1}^{j}$ to $\tilde{C}_{k+1}^{i,j}$. Just as before, \cite{wang2020persistent} shows that the dimension of the kernel of  $\mathcal{L}_{k}^{i,j}$ is the $k$-th persistent Betti number $\beta_{k}^{i,j}$.

Going back to the VR complexes introduced earlier, it follows from their definition that if $\eps\le\eps'$ then $S^{\eps}\subseteq S^{\eps'}$. Moreover, since there is only a finite number of data points, $S^{\eps}$ will remain unchanged for all $\eps>\text{diam}(\bm x)$. So, there is a maximal VR complex $S$ that contains all other VR complexes $S^{\eps}$, it's left to the reader to check that if the data consists of $n$ points then $S$ has exactly $2^n$ simplices. Choosing an increasing sequence of scales $0<\eps_1\le\eps_2\le\dots\eps_N$ yields a filtration of VR complexes $\emptyset \subseteq S^{\eps_{1}} \subseteq S^{\eps_{2}} \subseteq ... \subseteq S^{\eps_{N}}=S$.

\section{Quantum algorithm}
\label{sec:3}

% {\color{red}Maybe a short outline of how classical algorithms work to compare}

Quantum algorithms proceed by first mapping the simplices, simplicial complexes, and filtrations onto quantum states, and then implementing linear unitary operators acting on them in order to extract topological features through measurements of quantum states. All $2^n$ possible simplices can be mapped onto quantum states in an $n$-qubit Hilbert space enabling an exponential speedup.

Our quantum algorithm uses a quantum Random Access Memory (QRAM) to access (or estimate) data in quantum parallel and create quantum states that encode the VR complexes. Indeed, the $k-$simplices, which are fully connected sets of $k+1$ vertices, are encoded as quantum states over $n$ qubits with $k+1$ qubits corresponding to the vertices of the simplex in the state $|1\rangle$ and the rest of the $n-k-1$ qubits in the state $|0\rangle$. The boundary of a $k$-simplex can be obtained as a superposition of all the $(k-1)-$simplices that result from flipping one of the qubits in the state $|1\rangle$ to $|0\rangle$. Furthermore, the projection operators necessary to express the persistent combinatorial Laplacian can be implemented using an oracle based on Grover's search algorithm. Then, one may form a superposition of simplices and use a quantum phase estimation algorithm to extract information about the kernel of the persistent combinatorial Laplacian, thus obtaining the desired persistent Betti numbers.
It is important to note that the resources needed to encode the simplicial complex (number of qubits) are only polynomial in the number of data points $n$, while in the classical case, the resources needed grow exponentially with $n$.

%Our algorithm uses a quantum Random Access Memory (QRAM) to map classical data, such as positions of the data-points or pairwise distances, into a set of quantum states {\color{red} Cite 10}\cite{QRAM, giovannetti2008architectures}. It constructs simplicial complexes by encoding simplices as the basis states of a Hilbert space. Then it uses a quantum phase estimation algorithm {\color{red} Cite papers} to obtain information about the kernel of a linear operator, which is shown later in this paper to be the desired topological information.

\subsection{Encoding simplices and filtrations as quantum states}

%If the distances between points are already known, they can be stored in the qRAM and accessed in quantum parallel in order to construct the filtration. 

%We store simplices in a $n$-qubit register. A $k$-simplex $\s = [v_{i_1},\dots,v_{i_k}]$ is stored as $|\s\ra = |v_1\ra \otimes \dots \otimes |v_n\ra$, with 1s at the positions of its $k+1$ vertices $v_{i_1},\dots,v_{i_k}$ and 0s elsewhere. 
A $k$-simplex $\s = [v_{i_1},\dots,v_{i_k}]$ can be stored in a $n$-qubit register as $|\s\ra = |v_1\ra \otimes \dots \otimes |v_n\ra$, with 1s at the positions of its $k+1$ vertices $v_{i_1},\dots,v_{i_k}$ and 0s elsewhere. Let $S$ denote the collection of all possible states of $n$ qubits, then $S$ effectively encodes the maximal VR complex described in section \ref{sec:2} with all possible simplices that can be formed with $n$ vertices or data points. Moreover, $\HH$, the Hilbert space over $\Z_{3}$ with basis $S$ encodes the chain group of that complex.

We denote by $S_{k}$ the subset of states in $S$ encoding $k-$simplices, and by $\HH_{k}$ the corresponding subspace of $\HH$ which encodes the $k$-th chain group defined in \eqref{eqn_chain_group}. We can encode the order $k$ of a simplex $\s$ in a state $|k\rangle$ ($k=0,1,\dots,n-1$) by starting from $|0\rangle$ and performing permutations $0\to 1 \to \dots \to n-1 \to 0$, conditional upon the corresponding digit of $\s$ being 1. Thus, we perform $k$ permutations mapping $|0\rangle \to |k\rangle$. This can be implemented efficiently because the permutation is a 1-sparse matrix.

Similarly, we write $S^{\eps}$ for the subset of $S$ encoding simplices of diameter at most $\eps$, in other words the VR complex at scale $\eps$ as in Def. \ref{defn_vr}, and $\HH^{\eps}$ for the corresponding subspace of $\HH$. To encode the scale $\eps$ we need information on the data points that can be stored in quantum parallel in QRAM, if it is available, and accessed efficiently \cite{QRAM, giovannetti2008architectures}. For any $i,j = 1,2,\dots, n$, 
$\mbox{QRAM}|i\rangle |j\rangle |0\rangle=|i\rangle |j\rangle |d(i,j)\rangle$, where $d(i,j)$ is the distance between points $i$ and $j$.
% \begin{equation}
% 	\text{QRAM} |i\rangle |j\rangle |0\rangle = |i\rangle |j\rangle |d(i,j)\rangle
% \end{equation}
We introduce a register of qubits to record the parameter $\epsilon$ as $|\epsilon\rangle$. We need to know when $d(i,j) \le \epsilon$ to form a VR complex. This information will be stored in a qubit initially in the state $|0\rangle$, and flipped if the membership condition is satisfied. This is implemented with a unitary test that uses the qubit registers storing $d(i,j)$ and $\epsilon$ as controls to flip the last qubit,
\begin{equation}
	U_{\text{test}}^\epsilon |d(i,j)\rangle |\epsilon\rangle |0\rangle = |d(i,j)\rangle |\epsilon\rangle |a^\epsilon(i,j)\rangle \ , \ \ a^\epsilon(i,j) = \left\{ \begin{array}{ccc}
0 & , & d(i,j) > \epsilon \\ 1 & , & d(i,j) \le \epsilon
	\end{array} \right.
\end{equation}
Next, in order to know if $\s \in S^{\eps}$, we must check if $d(i,j) \le\epsilon$ for all $(i,j)$ pairs such that $v_i = v_j = 1$. To this end, we make $\mathcal{O} (k^2)$ calls to QRAM, where $k$ is the dimension of $\s$. For each pair $(i,j)$, we use $|\s\ra$ as control to call QRAM and apply the test provided $v_i = v_j = 1$,

\begin{equation}
\text{QRAM}^\dagger U_{\text{test}}^\eps \text{QRAM} |\s\ra |i\ra |j\ra |0\ra |\eps\ra |0\ra = |\s\ra |i\ra |j\ra |0\ra |\eps\ra |a^\eps (i,j)\ra 
\end{equation}
Membership of $\sigma$ in the VR complex $S^\eps$ is decided if for all $(i,j)$ we end up with $a^\eps (i,j) =1$. 
%Therefore, we ought to stop at the first flip of the last qubit. This can be done by making the flipping conditional on $s_k(i) = s_k(j) = a^\epsilon (i,j)= 1$. At the end, if the last qubit is $|0\rangle$ ($|1\rangle$), then $s_k\notin S_k^\epsilon$ ($s_k\in S_k^\epsilon$).
The above steps allow us to implement the quantum oracle
\begin{equation}\label{eq:4}
\mathcal{O}_k^\eps	|\s\ra|0\ra = |\s\ra |a_{\s}^\eps\ra \ , \ \ a_{\s}^\eps = \left\{ \begin{array}{ccc}
		0 & , & \s\notin S_k^\eps \\ 1 & , & \s\in S_k^\eps
	\end{array} \right.
\end{equation}
determining membership in $S_{k}^{\eps} = S_{k}\cap S^{\eps}$, which encodes the $k$-simplices in the $\eps$-complex.
The oracle in Eq. \eqref{eq:4} can be used for the implementation of the projection operators,
\be\label{eq:P} P_k^\epsilon =  \sum_{\s\in S_k^\epsilon} |\s\rangle\langle \s|, \ee
onto $\HH_k^\epsilon,$ the subspace of the Hilbert space $\HH_k$ %of all $\binom{n}{k+1}$ $k$-simplices,    
spanned by the simplices in $S_k^\epsilon$, which are needed to construct the persistent combinatorial Laplacian. To implement the projection $P_k^\eps$, we perform amplitude amplification  \cite{brassard2002quantum} based on Grover's search algorithm \cite{grover} (for details, see Appendix \ref{app:A}).
%that encodes the $k$-th persistent chain group at scale $\eps$. %and is generated by $n$-qubit states with exactly $k+1$ qubits in the state $|1\rangle$. The VR complex at scale $\epsilon$ is encoded as $\HH^\epsilon = \oplus_k \HH_k^\epsilon$.
% Vietoris Rips complex
%Let $\HH=\C^{2^n}$ denote the complete simplicial complex $S$ encoded as a Hilbert space where the basis vectors, as mentioned above, correspond to the $2^n$ simplices. We can view the set of $k$-simplices in $S$ as a subspace $\HH_k$ of the Hilbert space $\HH$ of dimension $\binom{n}{k+1}$ generated by the $n$-qubits with exactly $k+1$ qubits equal to 1. We can also consider , then  The projection operators
%onto the subspaces $\HH^\epsilon$ play a major role in the next part of the algorithm since they are used .

\subsection{Persistent Dirac operator}
%{\color{purple}Some of the notations here appear in Section 2 and we should call them from therein.}
The boundary operator $\partial_k$ mapping $\HH_{k}$ to $\HH_{k-1}$ as defined in \eqref{eqn_boundary_map} may be encoded using 1-qubit Pauli X gates, and expressed similarly by its action on the bases $S_k$ and $S_{k-1}$
\be\label{eq:b} \partial_k |\s_k\rangle := \sum_{l=0}^k (-)^lX_{i_l} |\s_k\rangle\ = \sum_{l=0}^k (-)^l |\s_{k-1} (l)\rangle  , \ee
where $|\s_{k-1} (l)\ra$ is the $(k-1)$-simplex obtained from $|\s\ra$ by flipping its $l$-th non-zero qubit. One easily deduces $\partial_{k-1} \partial_k = 0$.

For persistent homology we need to restrict the boundary operator \eqref{eq:b} as in the prelude to Eq.\ \eqref{eq:6},
%{\color{red} -- also a bit confusing: here we use $\eps, \eps'$, and before we used $i,j$.},
\be\label{eq:br} \tilde{\partial}_k^{\epsilon,\epsilon'} = P_{k-1}^\epsilon \partial_k P_k^{\epsilon'} \ , \ee
where the projections are defined in \eqref{eq:P}. Then the persistent combinatorial Laplacian $\mathcal{L}_{k}^{\eps,\eps'}$ is as in \eqref{eq:6} with $\K_i = S^{\eps}$ and $\K_j = S^{\eps'}$.
%, at two different scales $\epsilon,\epsilon'$, as needed for persistence analysis
% For $\epsilon' = \epsilon$, this definition reduces to the expression of the Laplacian, which does not provide the persistent Betti numbers, as introduced in \cite{lloyd2016quantum}.

%The persistent combinatorial Laplacian is given by
%\be\label{eq:pcl} \mathcal{L}^{\epsilon,\epsilon'} = \tilde{\partial}^{\epsilon,\epsilon\dagger} \tilde{\partial}^{\epsilon,\epsilon} + \tilde{\partial}^{\epsilon,\epsilon'} \tilde{\partial}^{\epsilon,\epsilon'\dagger}\ee
Next, we introduce the persistent Dirac operator which plays a central role in our quantum algorithm. For quantum computation, it is advantageous to work with an operator which is linear in the boundary map. Unlike the persistent combinatorial Laplacian, the persistent Dirac operator is linear in the boundary operator, and its square yields the persistent combinatorial Laplacian given in Eq. \eqref{eq:6}.
% Short explanation of topological data analysis
%Persistent homology requires  the consideration of linear maps that act on the space of simplices. For quantum computation, it is advantageous to work with linear operators. However, the Laplacian is quadratic and yet quantum computers require maps to be Hermitian or selfadjoint. To that end, we need to build a Hermitian operator that encodes the Laplacian.

We define the $k$-th persistent Dirac operator as the Hermitian operator
\be\label{eq:18} {B}^{\epsilon,\epsilon'}_k = \left( \begin{array}{ccc}  0 &  \tilde{\partial}_k^{\epsilon,\epsilon} & 0\\
  \tilde{\partial}_k^{\epsilon,\epsilon\dagger} & 0 & \tilde{\partial}_{k+1}^{\epsilon,\epsilon'} \\
  0 &\tilde{\partial}_{k+1}^{\epsilon,\epsilon'\dagger} & 0
\end{array}\right) - \xi \left( \begin{array}{ccc}  P^\eps_{k-1} &  0 & 0\\
  0 & -P^\eps_k & 0 \\
  0 &0 & P^{\eps'}_{k+1}
\end{array}\right), \ee
where $\xi\in\mathbb{R}$ is an arbitrary parameter. Its square is
%\be \left( {B}^{\epsilon,\epsilon'}_k \right)^2 = \left( \begin{array}{ccc}   %\tilde{\partial}_k^{\epsilon,\epsilon} \tilde{\partial}_k^{\epsilon,\epsilon\dagger} +\xi^2 %P^\eps_{k-1}  &  0 & \tilde{\partial}_k^{\epsilon,\epsilon} %\tilde{\partial}_{k+1}^{\epsilon,\epsilon'} \\
%  0 & \mathcal{L}_k^{\eps,\eps'} + \xi^2 P^\eps_k & 0 \\
%  \tilde{\partial}_{k+1}^{\epsilon,\epsilon'\dagger} %\tilde{\partial}_k^{\epsilon,\epsilon\dagger} &0 &  %\tilde{\partial}_{k+1}^{\epsilon,\epsilon'\dagger} %\tilde{\partial}_{k+1}^{\epsilon,\epsilon'} + \xi^2  P^{\eps'}_{k+1}
%\end{array}\right) \ee
%
\be \left( {B}^{\epsilon,\epsilon'}_k \right)^2 = \left( \begin{array}{ccc}   \tilde{\partial}_k^{\epsilon,\epsilon} \tilde{\partial}_k^{\epsilon,\epsilon\dagger} &  0 & \tilde{\partial}_k^{\epsilon,\epsilon} \tilde{\partial}_{k+1}^{\epsilon,\epsilon'} \\
  0 & \mathcal{L}_k^{\eps,\eps'} & 0 \\
  \tilde{\partial}_{k+1}^{\epsilon,\epsilon'\dagger} \tilde{\partial}_k^{\epsilon,\epsilon\dagger} &0 &  \tilde{\partial}_{k+1}^{\epsilon,\epsilon'\dagger} \tilde{\partial}_{k+1}^{\epsilon,\epsilon'}
\end{array}\right) +  \xi^2 \left( \begin{array}{ccc}    P^\eps_{k-1}  &  0 & 0 \\
  0 & P^\eps_k & 0 \\
  0 & 0 & P^{\eps'}_{k+1}
\end{array}\right) \ee
written in block form with one of the blocks being the $k$th persistent combinatorial Laplacian shifted by $\xi^2$. 

Let $\lambda$ be an eigenvalue of ${B}^{\epsilon,\epsilon'}_k$. Evidently, the corresponding eigenspace corresponds to an eigenspace of the Laplacian $\mathcal{L}_k^{\eps,\eps'}$ with eigenvalue
$ \gamma = \lambda^2 - \xi^2. $ Considering $\gamma = 0,$ one gets the the kernel of $\mathcal{L}_k^{\eps,\eps'},$  whose dimension is the $k$th persistent Betti number,
$ \beta_k^{\epsilon,\epsilon'} = |\ker \mathcal{L}^{\epsilon,\epsilon'}_k |.$
In other words, one is interested in the eigenspaces of ${B}^{\epsilon,\epsilon'}_k$ with eigenvalues $\lambda = \pm \xi$. For the persistent Betti numbers, we focus on the eigenspace with $\lambda = \xi$, because there is a one-one correspondence between its states and those in the null space of the Laplacian. Thus, although $\xi$ is arbitrary, we need to choose $\xi \ne 0$ in order to separate the eigenspaces with $\lambda = - \xi$ and $\lambda = \xi$. The choice $\xi =0$ leads to overcounting of the states in the null space of the Laplacian and an incorrect estimate of Betti numbers. 

To see this, consider the action on vectors of the form $(X_{k-1}^\epsilon, Y_k^\epsilon, Z_{k+1}^{\epsilon'} )^T$ which satisfy the constraints $P^\epsilon X_{k}^\epsilon =X_k^\eps$, $P^\epsilon Y_k^\epsilon = Y_k^\eps$, and $P^{\epsilon'} Z_{k}^{\epsilon'} =Z_k^{\eps'}$. 
For eigenvectors of $B^{\eps,\eps'}_k$ corresponding to eigenvalue $\lambda$, we obtain the system of equations
\begin{align}
  - {\xi} X_{k-1}^\epsilon + \tilde{\partial}_k^{\epsilon,\epsilon} Y_k^\epsilon &= \l X_{k-1}^\epsilon \nonumber\\
  {\xi} Y_k^\eps + \tilde{\partial}_k^{\epsilon,\epsilon\dagger} X_{k-1}^\epsilon + \tilde{\partial}_{k+1}^{\epsilon,\epsilon'} Z_{k+1}^{\epsilon'}  &= \l Y_k^\epsilon \nonumber\\
  - {\xi} Z_{k+1}^{\epsilon'} + \tilde{\partial}_{k+1}^{\epsilon,\epsilon'\dagger} Y_k^\epsilon &= \l Z_{k+1}^{\epsilon'} 
\end{align}
For $\lambda \ne -\xi$, we can use two of the equations to express $X_{k-1}^\epsilon$ and $Z_{k+1}^{\epsilon'}$ in terms of $Y_k^\eps$, and the second equation to show that $Y_k^\eps$ is an eigenvector of the Laplacian with eigenvalue $\gamma$. Conversely, given an eigenvector $Y_k^\eps$ of the Laplacian with eigenvalue $\gamma$, the vector $(X_{k-1}^\epsilon, Y_k^\epsilon, Z_{k+1}^{\epsilon'} )^T$ with  $X_{k-1}^\epsilon = \frac{1}{\lambda + \xi } \tilde{\partial}_k^{\epsilon,\epsilon} Y_k^\epsilon$ and $Z_{k+1}^{\epsilon'} = \frac{1}{\lambda + \xi } \tilde{\partial}_{k+1}^{\epsilon,\epsilon'\dagger} Y_k^\epsilon$ is an eigenvector of the Dirac operator with eigenvalue $\lambda$, demonstrating a one-one correspondence between the two eigenspaces. 
%
%where
%
%\be \mathcal{L}_{\epsilon,\epsilon'}^k  = \tilde{\partial}_k^{\epsilon,\epsilon\dagger} \tilde{\partial}_k^{\epsilon,\epsilon}   + \tilde{\partial}_{k+1}^{\epsilon,\epsilon'} \tilde{\partial}_{k+1}^{\epsilon,\epsilon'\dagger} \ee
%
%Moreover, since the two terms on the left are orthogonal, we can split the above into two equations, one for each component of $Z_k^\epsilon$:
%
%\begin{align}\label{eq:29}
%  \mathcal{L}_{\epsilon,\epsilon'}^k P_{k}^{\eps} Z_k^\epsilon & = \l(\l -\a) P_{k}^{\eps} Z_k^\epsilon \nonumber\\
%  (\l -\a) Q_{k}^{\eps} Z_k^\epsilon &= 0
%\end{align}

For $\l = -\xi$, we obtain eigenvectors with $Y_k^\eps$ in the kernel of the Laplacian. We need to separate them from the eigenvectors with $\l = {\xi}$ in order not to overcount the vectors in the null space of the Laplacian.

The persistent Dirac operator \eqref{eq:18} can be decomposed as
\be\label{eq:18a} {B}^{\epsilon,\epsilon'}_k = P_k^{\eps,\eps'} B_k P_k^{\eps,\eps'} \ , \ B_k = \left( \begin{array}{ccc}  -\xi I &  \partial_k & 0\\
  \partial_k^{\dagger} & \xi I & {\partial}_{k+1} \\
  0 & {\partial}_{k+1}^{\dagger} & -\xi I
\end{array}\right) \ , \ P_k^{\eps,\eps'} =  \left( \begin{array}{ccc}  P^\eps_{k-1} &  0 & 0\\
  0 & P^\eps_k & 0 \\
  0 &0 & P^{\eps'}_{k+1}
\end{array}\right), \ee
%To keep track of $k$-simplices, we also need to introduce a qubit label $|k\rangle$ and the ladder operator
%\be \bm{L} = \sum_k |k-1\rangle \langle k| \ee 
%mapping $|k\rangle$ to $|k-1\rangle$. It is easy to check that it is unitary ($\bm{L}^\dagger \bm{L} = I$).
%For $\eps' = \eps$, we consider the space spanned by $|k\rangle |\s\rangle$ with $\s \in S_k^\eps$, and define the Dirac operator
%\be B^\eps = P^\eps B P^\eps \ , \ \ B = \bm{L} \otimes \partial + \text{h.c.} \ee 
%where $\text{h.c.}$ denotes hermitian conjugate. By construction, $B^\eps$ is a hermitian operator whose square is the Laplacian: $(B^\eps)^2 = P^\eps (\partial^\dagger \partial + \partial \partial^\dagger ) P^\eps$, which is $\mathcal{L}^{\eps,\eps}$ when acting on $\s\in S_k^\eps$.
%If $\epsilon' \ne \eps$, we need to enlarge our space. 
To encode these $3\times 3$ block matrices, we introduce a qutrit label $|b\rangle$, with $b\in \{ -1,0,+1\}$, and consider the space spanned by $|b\rangle  |\s_{k+b}^b\rangle$, where $\s_{k+b}^b \in S_{k+b}^{\eps_b}$, with $\eps_b = \eps$ for $b=-1,0$ and $\eps_b = \eps'$ for $b= +1$. Then we may write
%We define the \textit{persistent} Dirac operator
%\be B^{\eps,\eps'} = P^{\eps,\eps'} B P^{\eps,\eps'} \ee 
%
%\noindent where we introduced projection operators in the enlarged space,
\be P_k^{\eps,\eps'} = \sum_{b\in \{ -1,0,+1 \} } |b\rangle\langle b| \otimes P_{k+b}^{\eps_b}  \ , \ee
and %the modified Dirac operator
\be\label{eq:B} B_k = \sum_{b\in \{ 0,+1 \} }  |b-1\rangle\langle b| \otimes \partial_{k+b}  + \text{h.c.} + \xi \sum_{b\in \{ -1,0,+1 \} } (-)^b |b\rangle\langle b|  \otimes I \ee
%with $\alpha$ an arbitrary parameter. The term proportional to $\alpha$ is added to move spurious solutions (due to the enlargement of space) outside of the kernel.
The projections $P_k^{\eps,\eps'}$ can be implemented using Grover's algorithm as outlined above and described in detail in Appendix \ref{app:A}, whereas the unrestricted Dirac operator $B_k$ can be encoded in a qubit register. For $\xi =1$, a qutrit suffices for the encoding of $B_k$, because all of its matrix elements are in the set $\{ -1,0,+1 \}$.

\subsection{Persistent Betti numbers}

As discussed earlier, the $k$th persistent Betti number %at levels $(\eps,\eps')$, 
$\beta_k^{\eps,\eps'}$ is the dimension of the eigenspace of the persistent Dirac operator $B_k^{\eps,\eps'}$ (Eq.\ \eqref{eq:18}) with eigenvalue $\lambda = \xi$. To compute it, we apply phase estimation (for details, see Appendix \ref{app:C}), and derive the probability distribution
\begin{equation}\label{eq:9}
	\mathcal{P} (p) = \frac{1}{\bm{N}} \sum_{\lambda} g_\lambda ( p) \ , \ \ g_\lambda (p) = \frac{1}{M^2} \frac{\sin^2 \pi l\lambda }{\sin^2 \frac{\pi(l\lambda -p)}{M}} \ ,
\end{equation}
where $l, M$ are arbitrary positive integers, $p=0,1,\dots, M-1$, and $\bm{N}$ is the number of eigenvalues $\lambda$ of $B_k^{\eps,\eps'}$. One easily checks that $\sum_{p=0}^{M-1} g_\lambda (p) = 1$ for any $\lambda$, therefore, $\sum_{p=0}^{M-1} \mathcal{P} (p) = 1$, as expected.

If one is interested in the spectrum of the persistent Dirac operator, $M$ and $l$ must be chosen in a way that  the spectrum is covered with the different values of $p$, i.e., each eigenvalue is approximated as $\lambda \approx \frac{p}{l}$. Then $g_\lambda (p)$ is strongly peaked at $p \approx l\lambda$. At the peak, $g_\lambda (p) = 1$. Therefore, the probability distribution $\mathcal{P} (p)$ has a peak at each eigenvalue of the persistent Dirac operator whose height is proportional to the dimension of the corresponding eigenspace. 

Since we are only interested in the eigenvalue $\lambda = \xi$, we only need to capture the peak at $p \approx l\xi$ and make sure it is resolved from other neighboring peaks. The persistent Betti number is obtained as $\beta_k^{\eps,\eps'} = \left. g_\xi (p) \right|_{p\approx l\xi}$.

\section{An application}
\label{sec:4}

%At this point you're probably wondering if this algorithm could run on a current quantum computer. Unfortunately, the answer is no. There are certain features of this algorithm that cannot yet be implemented on current quantum computers. Moreover, the size of current quantum computers is not large enough to be able to analyze a useful data set. What can be done, is 
This Section demonstrates how our algorithm computes persistence Betti numbers, which track topological features across different scales. This extends previous work in \cite{lloyd2016quantum} and \cite{siopsis2019quantum} where the proposed quantum algorithms calculated Betti numbers only without addressing persistence features. To do this, we apply our method to the data set suggested in  \cite{gunn2019review} and described below for the sake of completeness.

\begin{figure}[ht]
\begin{center}
\includegraphics[width=\textwidth]{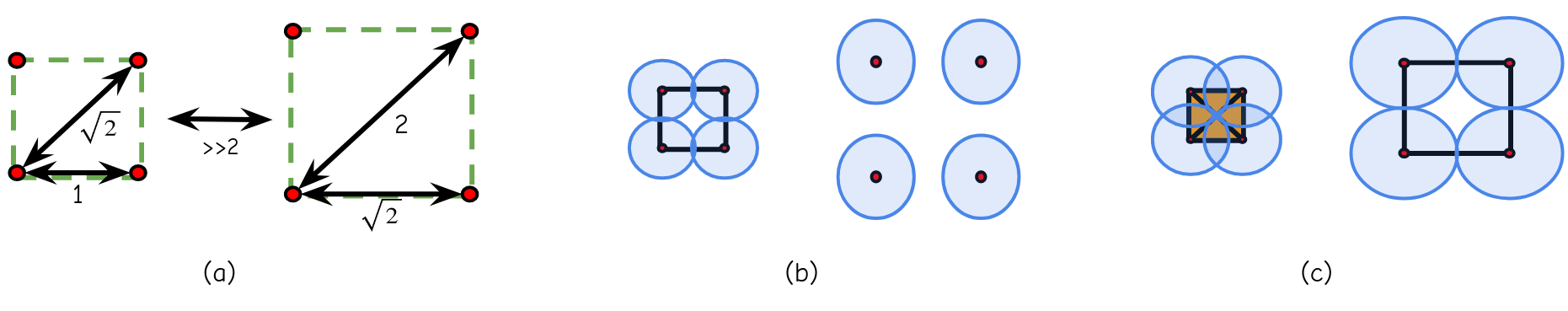} 

\end{center}
\caption{
\textit{(a)} The two squares example as appeared in \cite{gunn2019review}; \textit{(b)} Construction of the VR for $\epsilon_1$; \emph{(c)} VR for $\epsilon_2$} \label{fig:two_squares}
\end{figure}

Consider a point cloud of 8 points consisting of two well-separated squares, as in Figure \ref{fig:two_squares}. The smaller square has sides of length 1, while the larger square has sides of length $\sqrt{2}$. The distance between the two squares exceeds 2. It is easy to see (Figure \ref{fig:two_squares}) that the smaller square produces a loop in the VR complex at scale $1$ which disappears at scale $\sqrt{2}$, while at the same time the larger square produces a new loop.
It follows that for
$1 < \eps_{1} < \sqrt{2} < \eps_{2} < 2 $,
the one--dimensional persistent Betti numbers corresponding to the point cloud of Figure \ref{fig:two_squares} are
\be \beta_{1}^{\eps_1,\eps_1} = 1 \ , \ \beta_{1}^{\eps_2,\eps_2} = 1 \ , \ \beta_{1}^{\eps_1,\eps_2} = 0 \ . \ee
It should be pointed out that the algorithms proposed in \cite{lloyd2016quantum,siopsis2019quantum} can detect the number of loops in the VR complex at scales $\eps_1$ and $\eps_2$ by computing  $\beta_{1}^{\eps_1,\eps_1}$ and $\beta_{1}^{\eps_2,\eps_2}$, respectively. However these Betti numbers do not hold any persistence information. The algorithms in \cite{lloyd2016quantum,siopsis2019quantum} cannot calculate the persistence Betti number $\beta_{1}^{\eps_1,\eps_2}$ which holds the persistence information (number of loops present at scale $\eps_1$ that persist to scale $\eps_2$). Thus, these algorithms cannot track topological features across different scales.
Moreover, even though $\beta_{1}^{\eps_1,\eps_1} = \beta_{1}^{\eps_2,\eps_2}$, the persistence Betti number cannot be deduced from $\beta_{1}^{\eps_1,\eps_1}$ and $\beta_{1}^{\eps_2,\eps_2}$. This is because it indicates that the loops at scales across $\sqrt{2}$ are different, which is additional information to the existence of a loop encoded in the other two Betti numbers.

\begin{figure}[ht]
  \centering
  \subfigure[]{\includegraphics[width=0.3\textwidth]{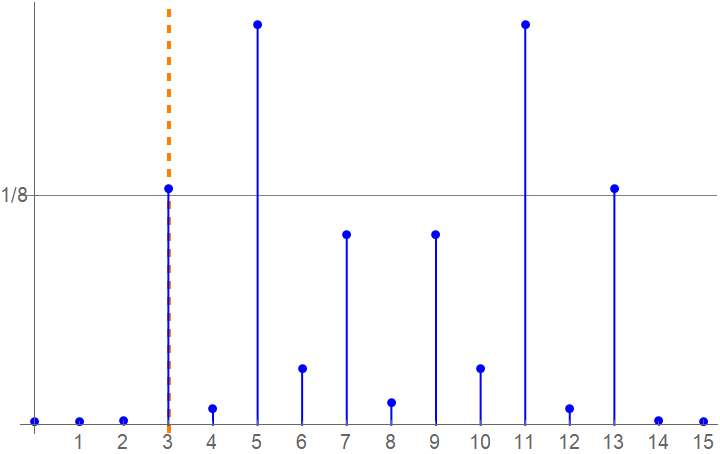}}
  \subfigure[]{\includegraphics[width=0.3\textwidth]{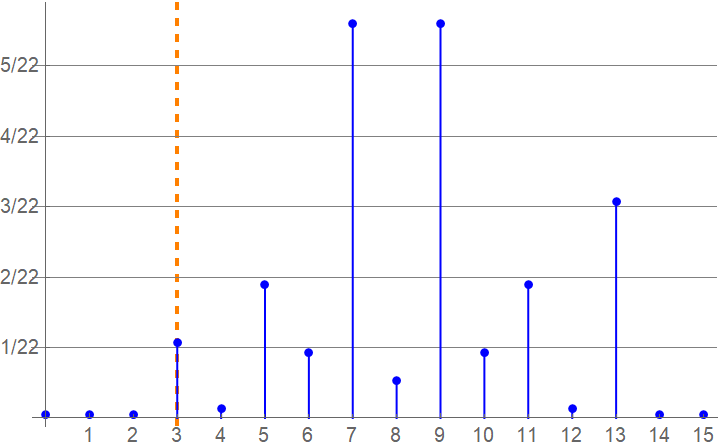}}
 \subfigure[]{\includegraphics[width=0.3\textwidth]{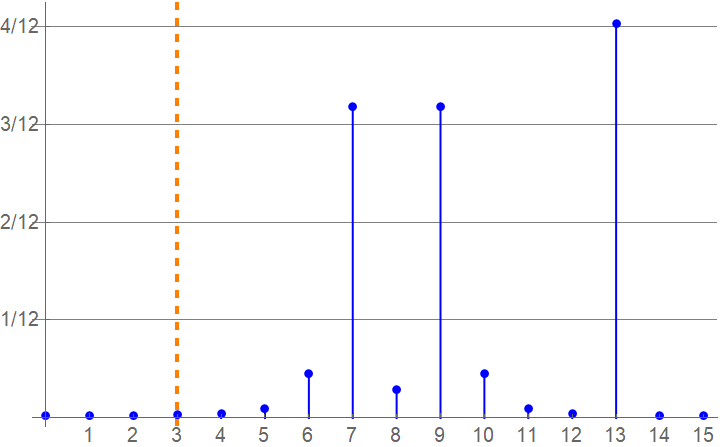}}
  \caption{Probability density corresponding to the persistent Dirac operator (a) $B_{1}^{\eps_1,\eps_1}$, (b) $B_{1}^{\eps_2,\eps_2}$, and (c) $B_{1}^{\eps_1,\eps_2}$, with $\xi = 1$, $l=3$, $M=16$. The heights at $p=3$, multiplied by the dimension of the Hilbert space (8, 22, and 12, respectively), yield the Betti numbers $\beta_1^{\eps_1,\eps_1} = \beta_1^{\eps_2,\eps_2} = 1$, and the persistent Betti number $\beta_1^{\eps_1,\eps_2} =0$.} \label{fig:2}
\end{figure}

%On the other hand, applying the quantum phase estimation algorithm to the corresponding persistent Dirac operators would yield a superposition of eigenvectors as in equation \eqref{eq:8a}.

%In this simple case, one can obtain the necessary boundary matrices by hand and use mathematical software to find the eigenvalues of the persistent Dirac operator, and then plot the corresponding probability distribution \eqref{eq:9}, as shown in \ref{fig:2}. In {\color{red} Cite figure a, b} we can see that the probability density corresponding to $B_{1}^{\eps_{1},\eps_{1}}$, $B_{1}^{\eps_{2},\eps_{2}}$  has a peak at XXX of height $1/N$ which agrees with our expectation of $\beta_{1}^{\eps_{1},\eps_{1}}=1$, $\beta_{1}^{\eps_{2},\eps_{2}}=1$. On the other hand, {\color{red} Cite figure c} presents the probability distribution corresponding to $B_{1}^{\eps_{1},\eps_{2}}$, this time there is no peak at XXX which correctly captures the persistence information $\beta_{1}^{\eps_{1},\eps_{2}}=0$.

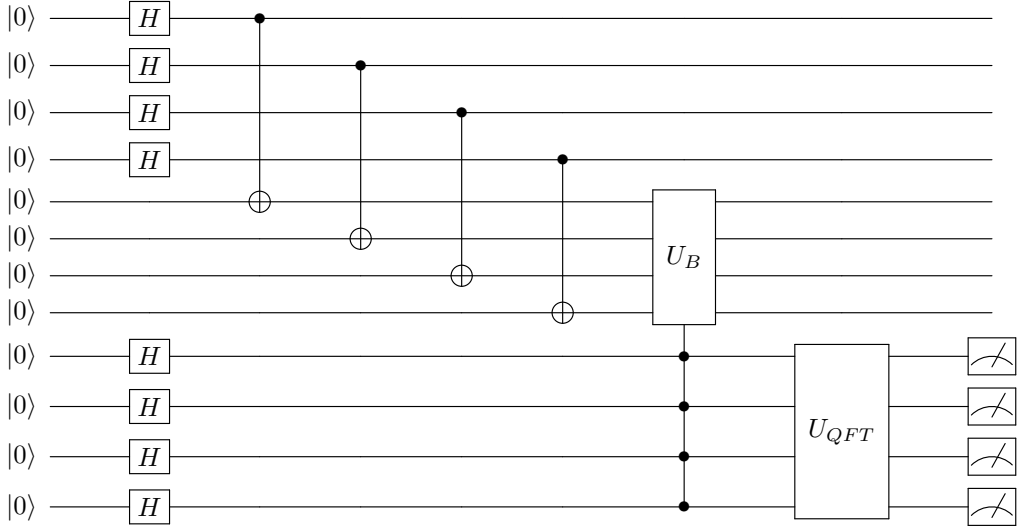
\begin{figure}[ht]
\[
    \Qcircuit @C=3em @R=.5em {
    \lstick{\ket{0}} & \gate{H}  & \ctrl{4}  & \qw & \qw & \qw & \qw & \qw & \qw \\
    \lstick{\ket{0}} & \gate{H} & \qw & \ctrl{4} & \qw &\qw &\qw &\qw & \qw \\
    \lstick{\ket{0}} & \gate{H} & \qw & \qw & \ctrl{4} & \qw & \qw &\qw &\qw \\
    \lstick{\ket{0}} & \gate{H} & \qw & \qw & \qw & \ctrl{4} & \qw & \qw &\qw \\
       \lstick{\ket{0}} & \qw  & \targ  & \qw & \qw & \qw & \multigate{3}{U_B} & \qw & \qw \\
    \lstick{\ket{0}} & \qw & \qw & \targ & \qw & \qw &\ghost{U_B} &\qw &\qw \\
    \lstick{\ket{0}} & \qw & \qw & \qw & \targ & \qw & \ghost{U_B} & \qw   &\qw \\
    \lstick{\ket{0}} & \qw & \qw & \qw & \qw & \targ & \ghost{U_B} & \qw   &\qw \\
   \lstick{\ket{0}} & \gate{H} & \qw & \qw  & \qw & \qw & \ctrl{-1} & \multigate{3}{U_{QFT}} & \meter \\
    \lstick{\ket{0}} & \gate{H} & \qw & \qw & \qw & \qw &\ctrl{-1} &\ghost{U_{QFT}} &\meter\\
    \lstick{\ket{0}} & \gate{H} & \qw & \qw & \qw &\qw & \ctrl{-1} & \ghost{U_{QFT}} & \meter \\
\lstick{\ket{0}} & \gate{H} & \qw & \qw & \qw & \qw & \ctrl{-1} & \ghost{U_{QFT}} & \meter
    } 
\]   \caption{Quantum circuit for the calculation of Betti number $\beta_1^{\eps_1,\eps_2}$.}\label{fig:qc}
\end{figure}

Figure \ref{fig:qc} shows the quantum circuit for the calculation of Betti number $\beta_1^{\eps_1,\eps_2}$. The Dirac operator $B_1^{\eps_1,\eps_2}$ acts on an 12-dimensional Hilbert space. For $\xi =1$ are $\lambda = -1, \pm\sqrt{5}$, each of degeneracy 4. Notice that $1$ is not an eigenvalue, therefore $\beta_1^{\eps_1,\eps_2} =0$. This is confirmed by the quantum algorithm depicted in Figure \ref{fig:qc} as explained below.

A register of 4 qubits initially in the state $|0000\rangle_1$ is brought into the state $\frac{1}{4} \sum_{x=0}^{15} |x\rangle_1$, by acting with the Hadamard matrix on each. Then we use them as control to apply CNOT on each of 4 qubits in an additional register in the state $|0000\rangle_2$, thus entangling them to the state $\frac{1}{4} \sum_{x=0}^{15} |x\rangle |x\rangle_2$. We introduce a third register of 4 qubits (choosing $M=16$) in the state $|0000\rangle_R$ and act on each with the Hadamard gate to bring them into the state $\frac{1}{4} \sum_{y=0}^{15} |y\rangle_R$. We then use them as control to act on register 2 with the exponential Dirac operator (see Appendix \ref{app:C} for details). Finally, we measure all 4 qubits in the register $R$. The result is the probability distribution $\mathcal{P} (p)$, where $p=0,1,\dots, 15$ (Eq.\ \eqref{eq:9}) depicted in Figure \ref{fig:2}(c). With the choice $l=3$, $M=16$, a measurement of the register of 4 qubits yields no peak at $p=3$ %of height $\mathcal{P} (3) = \frac{1}{8}$ 
, showing that $\beta_1^{\eps_1,\eps_2} =0$.

The persistent Betti number $\beta_1^{\eps_1,\eps_1}$ is calculated using an eight-dimensional Hilbert space. The eigenvalues of the persistent Dirac operator for $\xi =1$ are $\pm 1, \pm \sqrt{3}, \pm\sqrt{5}$. Two of the eigenvalues ($\pm\sqrt{3}$) are degenerate with multiplicity 2. We are interested in the multiplicity of the eigenvalue $\lambda = \xi = 1$, which is shown to be 1 by the peak at $p=3$ of height $1/8$ (see Fig.\ref{fig:2}(a)).  The closest eigenvalue to $\lambda =1$ is $\lambda = \sqrt{3}$ which is near $p=5$. Thus, the peak at the nearest eigenvalue is well separated from the one of interest (in Figure \ref{fig:2}(a), one can see a dip at $p=4$, and the height $\mathcal{P} (5) \approx \frac{2}{8}$, confirming the double degeneracy of the eigenvalue $\lambda = \sqrt{3}$).The quantum circuit flows as the one in Figure \ref{fig:qc} but with registers 1 and 2 having 3 qubits each. %The probability distribution showing that the persistent Betti number vanishes, even though $\beta_1^{\eps_1,\eps_1}$ and $\beta_1^{\eps_2,\eps_2}$ do not, is shown in Figure \ref{fig:2}(c).

The calculation of Betti number $\beta_1^{\eps_2,\eps_2}$ proceeds similarly. The Hilbert space is 22-dimensional and for $\xi =1$, the eigenvalues are the same as for $\beta_1^{\eps_1,\eps_1}$, but with degeneracies 1, 3, 2, 7
for $\lambda = 1, -1, \pm\sqrt{3}, \pm\sqrt{5}$, respectively. The distance from the eigenvalue of interest ($\lambda =1$) to its closest one ($\lambda = \sqrt{3}$) is same as before, therefore we can choose $l=3, M=16$, again. The quantum circuit is similar to the one in Figure \ref{fig:qc} except that registers 1 and 2 need 5 qubits each. The resulting probability distribution is depicted in Figure \ref{fig:2}(b) showing that $\beta_1^{\eps_2,\eps_2} =1$.

\section{Conclusion}
\label{sec:5}

% What does our algorithm provide?
Our work established a quantum persistent homology algorithm that detects and computes the topological features of point cloud data as their resolution changes. Our method considered the persistent Laplacian operator, generalizing the Laplacian operator discussed in \cite{lloyd2016quantum, siopsis2019quantum}. We provided an implementation on the challenging problem of two particular point cloud squares proposed in \cite{gunn2019review} as an example in which persistent Betti numbers cannot be deduced from Betti numbers. Classical algorithms for persistent homology (e.g., see \cite{zomorodian2005computing}) typically need $\mathcal{O}(2^{3n})$ operations to diagonalize a $2^{n}\times 2^{n}$ boundary matrix and obtain the topological information. Our quantum algorithm estimates the same information after a number of operations which is only polynomial in the number of data points $n$. Moreover, our algorithm can encode the full simplicial complex using $n$ qubits, while a classical computer needs $2^n$ bits for the same process.
Last, our algorithm needs a small number of measurements to obtain reliable results, and thus, its application to scientific data including signal processing or image analysis problems is advantageous.

% How is it better/different to the existing ones?

% On the other hand, we have shown that this algorithm can indeed track the topological features across different scales, unlike the previous quantum algorithms on the subject .

% {\color{red} Anything else?}
% % What are its weak points? What should we try to fix in the future?

% %Accuracy issues when spectral gap is small.

% %Does this run in current computers? If so how big of a data set could it analyze?

\acknowledgments
Research supported by the National Science Foundation award DMS-2012609.
G.\  Siopsis  also acknowledges the Army Research Office award W911NF-19-1-0397, and the National Science Foundation award OMA-1937008.

%\printbibliography
\bibliographystyle{unsrt}
\bibliography{references}

\begin{thebibliography}{10}

\bibitem{carlsson}
Gunnar Carlsson.
\newblock Topology and data.
\newblock {\em Bulletin of the American Mathematical Society}, 46(2):255--308,
  2009.

\bibitem{carlsson_basics}
Erik Carlsson, Gunnar Carlsson, and Vin~De Silva.
\newblock An algebraic topological method for feature identification.
\newblock {\em International Journal of Computational Geometry},
  16(4):291--314, 2006.

\bibitem{CompyTopo}
Herbert Edelsbrunner and John Harer.
\newblock {\em Computational topology: an introduction}.
\newblock American Mathematical Society, 2010.

\bibitem{vasudevan2013}
Ramanarayan Vasudevan, Aaron Ames, and Ruzena Bajcsy.
\newblock Persistent homology for automatic determination of human-data based
  cost of bipedal walking.
\newblock {\em Nonlinear Analysis: Hybrid Systems}, 7(1):101--115, 2013.

\bibitem{kusano2016}
Genki Kusano, Kenji Fukumizu, and Yasuaki Hiraoka.
\newblock Persistence weighted {G}aussian kernel for topological data analysis.
\newblock {\em arXiv:1601.01741}, 2016.

\bibitem{edelsbrunner2013}
Herbert Edelsbrunner.
\newblock Persistent homology in image processing.
\newblock In {\em International Workshop on Graph-Based Representations in
  Pattern Recognition}, pages 182--183. Springer, 2013.

\bibitem{chung2015}
Moo Chung, Jamie Hanson, Jieping Ye, Richard Davidson, and Seth Pollak.
\newblock Persistent homology in sparse regression and its application to brain
  morphometry.
\newblock {\em IEEE transactions on medical imaging}, 34(9):1928--1939, 2015.

\bibitem{munch2013}
Elizabeth Munch.
\newblock {\em Applications of persistent homology to time varying systems}.
\newblock PhD thesis, Duke University, 2013.

\bibitem{munch2016}
Firas~A Khasawneh and Elizabeth Munch.
\newblock Chatter detection in turning using persistent homology.
\newblock {\em Mechanical Systems and Signal Processing}, 70:527--541, 2016.

\bibitem{tda_action}
Vinay Venkataraman, Karthikeyan~Natesan Ramamurthy, and Pavan Turaga.
\newblock Persistent homology of attractors for action recognition.
\newblock {\em arXiv:1603.05310}, 2016.

\bibitem{tda_number}
Aaron Adcock, Erik Carlsson, and Gunnar Carlsson.
\newblock The ring of algebraic functions on persistence bar codes.
\newblock {\em Homology, Homotopy and Applications}, 18(1):381--402, 2016.

\bibitem{tda_wheeze}
Saba Emrani, Thanos Gentimis, and Hamid Krim.
\newblock Persistent homology of delay embeddings and its application to wheeze
  detection.
\newblock {\em IEEE Signal Processing Letters}, 21(4):459--463, 2014.

\bibitem{tda_clustering2015}
C{\'a}ssio~MM Pereira and Rodrigo~F de~Mello.
\newblock Persistent homology for time series and spatial data clustering.
\newblock {\em Expert Systems with Applications}, 42(15):6026--6038, 2015.

\bibitem{gunnarcancer}
Monica Nicolau, Arnold~J Levine, and Gunnar Carlsson.
\newblock Topology based data analysis identifies a subgroup of breast cancers
  with a unique mutational profile and excellent survival.
\newblock {\em Proceedings of the National Academy of Sciences},
  108(17):7265--7270, 2011.

\bibitem{HodgeCycle}
Joshua Mike and Vasileios Maroulas.
\newblock Combinatorial hodge theory for equitable kidney paired donation.
\newblock {\em Foundations of Data Science}, 1(1):87--101, 2019.

\bibitem{CPD_me}
Joshua Mike, Colin~D Sumrall, Vasileios Maroulas, and Fernando Schwartz.
\newblock Nonlandmark classification in paleobiology: computational geometry as
  a tool for species discrimination.
\newblock {\em Paleobiology}, pages 1--11, 2016.

\bibitem{Maroulas2020}
Vasileios Maroulas, Farzana Nasrin, and Christopher Oballe.
\newblock A bayesian framework for persistent homology.
\newblock {\em SIAM Journal on Mathematics of Data Science}, 2(1):48--74, 2020.

\bibitem{Townsend2020}
J.~Townsend, C.P. Micucci, J.H. Hymel, V.~Maroulas, and K.~D. Vogiatzis.
\newblock Representation of molecular structures with persistent homology for
  machine learning applications in chemistry.
\newblock {\em Nat Commun}, 11:3230, 2020.

\bibitem{Maroulas2021}
Vasileios Maroulas, Cassie~Putman Micucci, and Farzana Nasrin.
\newblock {Bayesian Topological Learning for Classifying the Structure of
  Biological Networks}.
\newblock {\em Bayesian Analysis}, pages 1 -- 26, 2021.

\bibitem{tda_windows}
Jose~A Perea and John Harer.
\newblock Sliding windows and persistence: An application of topological
  methods to signal analysis.
\newblock {\em Foundations of Computational Mathematics}, 15(3):799--838, 2015.

\bibitem{tda_tracking}
David Rouse, Adam Watkins, David Porter, John Harer, Paul Bendich, Nate Strawn,
  Elizabeth Munch, Jonathan DeSena, Jesse Clarke, and Jeffrey Gilbert.
\newblock Feature-aided multiple hypothesis tracking using topological and
  statistical behavior classifiers.
\newblock In {\em SPIE Defense+ Security}, pages 94740L--94740L. International
  Society for Optics and Photonics, 2015.

\bibitem{GdS06}
Vin De~Silva and Robert Ghrist.
\newblock Coordinate-free coverage in sensor networks with controlled
  boundaries via homology.
\newblock {\em The International Journal of Robotics Research},
  25(12):1205--1222, 2006.

\bibitem{GdS07}
Vin De~Silva and Robert Ghrist.
\newblock Homological sensor networks.
\newblock {\em Notices of the American mathematical society}, 54(1), 2007.

\bibitem{Ghrist12}
Pawel Dlotko, Robert Ghrist, Mateusz Juda, and Marian Mrozek.
\newblock Distributed computation of coverage in sensor networks by homological
  methods.
\newblock {\em Applicable Algebra in Engineering, Communication and Computing},
  23(1-2):29--58, 2012.

\bibitem{CdS1}
Gunnar Carlsson and Vin De~Silva.
\newblock Zigzag persistence.
\newblock {\em Foundations of computational mathematics}, 10(4):367--405, 2010.

\bibitem{CdS2}
Gunnar Carlsson, Vin De~Silva, and Dmitriy Morozov.
\newblock Zigzag persistent homology and real-valued functions.
\newblock In {\em Proceedings of the twenty-fifth annual symposium on
  Computational geometry}, pages 247--256. ACM, 2009.

\bibitem{persissensor}
Vin De~Silva and Robert Ghrist.
\newblock Coverage in sensor networks via persistent homology.
\newblock {\em Algebraic \& Geometric Topology}, 7(1):339--358, 2007.

\bibitem{jholes}
Jacopo Binchi, Emanuela Merelli, Matteo Rucco, Giovanni Petri, and Francesco
  Vaccarino.
\newblock jholes: A tool for understanding biological complex networks via
  clique weight rank persistent homology.
\newblock {\em Electronic Notes in Theoretical Computer Science}, 306:5--18,
  2014.

\bibitem{fullerene}
Kelin Xia, Xin Feng, Yiying Tong, and Guo~Wei Wei.
\newblock Persistent homology for the quantitative prediction of fullerene
  stability.
\newblock {\em Journal of computational chemistry}, 36(6):408--422, 2015.

\bibitem{tda_signal}
Mijail Guillemard and Armin Iske.
\newblock Signal filtering and persistent homology: an illustrative example.
\newblock {\em Proc. Sampling Theory and Applications (SampTA’11)}, 2011.

\bibitem{persis_brain}
Paul Bendich, James~Stephen Marron, Ezra Miller, Alex Pieloch, and Sean
  Skwerer.
\newblock Persistent homology analysis of brain artery trees.
\newblock {\em The Annals of Applied Statistics}, 10(1):198--218, 2016.

\bibitem{MaMa16}
Andrew Marchese and Vasileios Maroulas.
\newblock Topological learning for acoustic signal identification.
\newblock In {\em Information Fusion (FUSION), 2016 19th International
  Conference on}, pages 1377--1381. ISIF, 2016.

\bibitem{Marchese2018}
A.~Marchese and V.~Maroulas.
\newblock Signal classification with a point process distance on the space of
  persistence diagrams.
\newblock {\em Advances in Data Analysis and Classification}, 12(3):657--682,
  2018.

\bibitem{tda_timeseries}
Lee~M Seversky, Shelby Davis, and Matthew Berger.
\newblock On time-series topological data analysis: New data and opportunities.
\newblock In {\em Proceedings of the IEEE Conference on Computer Vision and
  Pattern Recognition Workshops}, pages 59--67, 2016.

\bibitem{tda_windowgenes}
Jose~A Perea, Anastasia Deckard, Steve~B Haase, and John Harer.
\newblock Sw1pers: Sliding windows and 1-persistence scoring; discovering
  periodicity in gene expression time series data.
\newblock {\em BMC bioinformatics}, 16(1):257, 2015.

\bibitem{dionysus}
Dmitriy Morozov.
\newblock Dionysus, a {C}++ library for computing persistent homology, 2007.

\bibitem{ripser}
Ulrich Bauer.
\newblock Ripser, 2015.
\newblock https://github.com/Ripser/ripser.

\bibitem{persistwist}
Chao Chen and Michael Kerber.
\newblock Persistent homology computation with a twist.
\newblock In {\em Proceedings 27th European Workshop on Computational
  Geometry}, volume~11, 2011.

\bibitem{lloyd2016quantum}
Seth Lloyd, Silvano Garnerone, and Paolo Zanardi.
\newblock Quantum algorithms for topological and geometric analysis of data.
\newblock {\em Nature communications}, 7(1):1--7, 2016.

\bibitem{siopsis2019quantum}
George Siopsis.
\newblock Quantum topological data analysis with continuous variables.
\newblock {\em Foundations of Data Science}, 4(1):419--431, 2019.

\bibitem{gunn2019review}
Sam Gunn and Niels Kornerup.
\newblock Review of a quantum algorithm for betti numbers.
\newblock {\em arXiv preprint arXiv:1906.07673}, 2019.

\bibitem{hayakawa2021quantum}
Ryu Hayakawa.
\newblock Quantum algorithm for persistent betti numbers and topological data
  analysis.
\newblock {\em arXiv preprint arXiv:2111.00433}, 2021.

\bibitem{Huang2018}
H.~Huang and \textit{et al.}
\newblock Demonstration of topological data analysis on a quantum processor.
\newblock {\em Optica}, 5:193, 2018.

\bibitem{1Qb}
R.~Dridi and H.~Alghassi.
\newblock Homology computation of large point clouds using quantum annealing.
\newblock {\em arXiv:1512.09328}, 2016.

\bibitem{wie2017quantum}
Chu~Ryang Wie.
\newblock A quantum circuit to construct all maximal cliques using grover
  search algorithm.
\newblock {\em arXiv preprint arXiv:1711.06146}, 2017.

\bibitem{wang2020persistent}
Rui Wang, Duc~Duy Nguyen, and Guo-Wei Wei.
\newblock Persistent spectral graph.
\newblock {\em International journal for numerical methods in biomedical
  engineering}, 36(9):e3376, 2020.

\bibitem{QRAM}
Vittorio Giovannetti, Seth Lloyd, and Lorenzo Maccone.
\newblock Quantum random access memory.
\newblock {\em Phys. Rev. Lett.}, 100:160501, Apr 2008.

\bibitem{giovannetti2008architectures}
Vittorio Giovannetti, Seth Lloyd, and Lorenzo Maccone.
\newblock Architectures for a quantum random access memory.
\newblock {\em Physical Review A}, 78(5):052310, 2008.

\bibitem{brassard2002quantum}
Gilles Brassard, Peter Hoyer, Michele Mosca, and Alain Tapp.
\newblock Quantum amplitude amplification and estimation.
\newblock {\em Contemporary Mathematics}, 305:53--74, 2002.

\bibitem{grover}
Lov~K. Grover.
\newblock Quantum computers can search rapidly by using almost any
  transformation.
\newblock {\em Phys. Rev. Lett.}, 80:4329--4332, May 1998.

\bibitem{zomorodian2005computing}
Afra Zomorodian and Gunnar Carlsson.
\newblock Computing persistent homology.
\newblock {\em Discrete \& Computational Geometry}, 33(2):249--274, 2005.

\bibitem{rebentrost2018quantum}
Patrick Rebentrost, Adrian Steffens, Iman Marvian, and Seth Lloyd.
\newblock Quantum singular-value decomposition of nonsparse low-rank matrices.
\newblock {\em Physical review A}, 97(1):012327, 2018.

\end{thebibliography}

\appendix

\section{Grover's algorithm}\label{app:A}

Here we review the salient features of amplitude amplification \cite{brassard2002quantum} and Grover's search algorithm \cite{grover} which are needed for the implementation of the projection $P_k^\eps$ (Eq.\ \eqref{eq:P}), for completeness.

Let $|\Psi_k\rangle\in\HH_{k}$ be a state in the span of the $k$-simplex states. We wish to construct the normalized projected state $|\Psi_k^\eps\rangle = \frac{P^\eps |\Psi_k\rangle}{\norm{P^\eps |\Psi_k\rangle}}\in\HH_{k}^{\eps}$, assuming that it exists. To this end, we introduce the unitary operator
\be U_G = -U_{\Psi_k} U^\eps \ , \ \ \text{with } \ \ U_{\Psi_k} = I - 2 |\Psi_k\rangle\langle \Psi_k | \ \text{and} \ \ U^\eps = I - 2P^\eps \ee
Splitting $|\Psi_k\rangle$ as
\be |\Psi_k\rangle = \sin\theta |\Psi_k^\eps\rangle + \cos\theta |\Bar{\Psi}_k^\eps \rangle \ , \ \ \sin\theta = \norm{P^\eps |\Psi_k\rangle} \ , \ee 
%{\color{red}Is $\norm{P^\eps |\Psi_k\rangle}$ equal to number of simplices (in the representation of $|\Psi_k\ra$) in $S_k^\eps$ over the number of simplices in $S_k$?}
where $P^\eps |\Psi_k^\eps\rangle = |\Psi_k^\eps\rangle$ and $P^\eps |\Bar{\Psi}_k^\eps\rangle = 0$, we can think of $|\Psi_k\rangle$ as the vector $\left( \begin{array}{c}
     \sin\theta  \\
     \cos\theta
\end{array} \right)$ in the two-dimensional space spanned by $\{ \Psi_k^\eps , \Bar{\Psi}_k^\eps \}$. Then $U_G$ acts as a rotation by an angle $2\theta$ and can be represented as $U_G = \left( \begin{array}{cc}
\cos 2\theta     & \sin 2\theta \\
-\sin 2\theta     & \cos 2\theta
\end{array} \right)$. 
Applying it $K$ times, we obtain the state
\be U_G^K |\Psi_k\rangle = \left( \begin{array}{c}
     \sin (2K+1)\theta  \\
     \cos (2K+1)\theta
\end{array} \right) \ee
This is close to the desired state for $(2K+1)\theta \approx \frac{\pi}{2}$. Therefore, the number of Grover steps needed is $K = \lfloor \frac{\pi}{4\theta} \rfloor$.
%{\color{red} Is $K$ then $\mathcal{O}\LP\LP \begin{array}{c} n \\ k+1 \end{array}\RP\RP$?}

\section{Implementation of an exponential operator}\label{app:B}

Here we review the construction of the exponential operator $e^{it {B}_k^{\epsilon,\epsilon'}}$ \cite{rebentrost2018quantum} which is needed for phase estimation (Eq.\ \eqref{eq:expC}).
We start by constructing the $\text{SWAP}_{{B}}$ operator from the Dirac matrix $B_k^{\epsilon,\epsilon'}$ (Eq.\ \eqref{eq:18a}),
%which is then used for the implementation of 
%
%From the persistent Dirac operator $\tilde{C}_{\epsilon,\epsilon'}$, we construct the SWAP-$\tilde{C}$ matrix
%
\begin{equation}
	\mathcal{S} \equiv \text{SWAP}_{{B}} = \sum_{x,y} {B}_k^{\epsilon,\epsilon'}(x,y) |y\rangle\langle x| \otimes |x\rangle \langle y| \ , \ \ {B}_k^{\epsilon,\epsilon'}(x,y) = \langle x |{B}_k^{\epsilon,\epsilon'} |y\rangle  \ ,
\end{equation}
where we have used labels $x,y$ to collectively represent $\{ b, \sigma_{k+b}^b \}$, where $b\in  \{ 0, \pm 1 \}$ and $\sigma_{k+b}^b \in S_{k+b}^{\eps_b}$, with $\eps_b = \eps$ for $b=-1,0$ and $\eps_b = \eps'$ for $b=+1$. With the choice $\xi = 1$, all matrix elements ${B}_k^{\epsilon,\epsilon'}(x,y)\in  \{ 0, \pm 1 \}$.

Then we construct the exponential $\text{SWAP}_{{B}}$ operator $e^{i\Delta t \mathcal{S}}$, which can be done efficiently because $\mathcal{S}$ is a one-sparse matrix.

Next, we act on the state $|\bm{s}\rangle \otimes |\Psi_k^{\eps,\eps'}\rangle$, where $|\bm{s}\rangle$ is the uniform state
\be\label{eq:s} |\bm{s} \rangle \propto P^{\epsilon,\eps'} \sum_{x} |x\rangle \ee
and $|\Psi_k^{\eps,\eps'}\rangle$ is an arbitrary state in the subspace on which $B_k^{\eps,\eps'}$ acts obeying $P^{\eps,\eps'} |\Psi_k^{\eps,\eps'}\rangle = |\Psi_k^{\eps,\eps'}\rangle$. They can be constructed using Grover's algorithm, as outlined in Appendix \ref{app:A}. After tracing over the space in which $|\bm{s}\rangle$ lives, we obtain
\be
\text{tr}_1 \left[ e^{-i \Delta t \mathcal{S}} |\bm{s}\rangle \langle \bm{s}|\otimes |\Psi_k^{\eps,\eps'}\rangle\langle\Psi_k^{\eps,\eps'} | e^{i\Delta t\mathcal{S}} \right]	= e^{-i {B}_k^{\eps,\eps'} \Delta t/\bm{N}} |\Psi_k^{\eps,\eps'}\rangle\langle\Psi_k^{\eps,\eps'} | e^{i {B}_k^{\eps,\eps'} \Delta t/\bm{N}}  + \mathcal{O} (\Delta t^2)
\ee
where $\bm{N}$ is the number of basis vectors comprising the state $\ket{\bm{s}}$ in \eqref{eq:s}. Then we apply Grover's algorithm to implement the projection $P^{\eps,\eps'}$ and obtain the state
\be e^{-i {B}_k^{\eps,\eps'} \Delta t/\bm{N}} |\Psi_k^{\eps,\eps'}\rangle \ee 
up to second order in $\Delta t$.

The desired state $e^{-it {B}_k^{\eps,\eps'} } |\Psi_k^{\eps,\eps'}\rangle$ for finite $t$ can be obtained by repeating the above construction as many times as needed.

\section{Phase estimation}\label{app:C}

% Phase estimation
Here we outline the steps of the phase estimation subroutine which plays a major role in our quantum algorithm leading to the probability distribution $\mathcal{P} (p)$, where $p=0,1,\dots,M-1$ and $M$ is an arbitrarily chosen integer (Eq.\ \eqref{eq:9}).

We start with the state \eqref{eq:s} which can be expanded in the computational basis as
\be |\bm{s}\rangle = \frac{1}{\sqrt{\bm{N}}} \sum_{x=0}^{\bm{N} -1} |x\rangle \ee
where $x$ collectively represents $\{ b, \sigma_{k+b}^b \}$, as detailed in Appendix \ref{app:B}.

Next, we copy the basis states to create the maximally entangled state
\begin{equation}
	|\tilde{\bm{s}}\rangle_{12} = \frac{1}{\sqrt{\bm{N}}} \sum_{x=0}^{\bm{N}-1} |x\rangle_1 |x\rangle_2
\end{equation}
Let the eigenvalue problem of the persistent Dirac operator $B^{\eps,\eps'}_k$ (Eq.\ \eqref{eq:18a}) be
\begin{equation}
	{B}^{\eps,\eps'}_k |\lambda\rangle = \lambda |\lambda\rangle
\end{equation}
By considering components, using the completeness of the set of eigenvectors, and the fact that they are real, it is easy to see that
\begin{equation}
	|\tilde{\bm{s}}\rangle_{12} = \frac{1}{\sqrt{\bm{N}}} \sum_{\lambda} |\lambda\rangle_1 |\lambda\rangle_2
\end{equation}
Next, we introduce a register of qubits $R$ in the state
\begin{equation}
	|R\rangle_R = \frac{1}{\sqrt{M}} \sum_{y=0}^{M-1} |y\rangle_R
\end{equation}
and apply the operator
\begin{equation}\label{eq:expC}
	U = e^{2\pi ily {B}^{\eps,\eps'}_k/M}
\end{equation}
entangling the registers 2 and $R$, where $M$ and $l$ are positive integers that can be adjusted at will. For details of the implementation of \eqref{eq:expC}, see Appendix \ref{app:B}. We obtain
\begin{equation}
	U|\tilde{\bm{s}}\rangle_{12}|R\rangle_R = \frac{1}{\sqrt{\bm{N}M}} \sum_{\lambda}\sum_{y=0}^{M-1} e^{2\pi i ly\lambda/M} |\lambda\rangle_1 |\lambda\rangle_2 |y\rangle_R
\end{equation}
Finally, we perform the quantum Fourier transform on the register $R$,
\begin{equation} U_{QFT}|y\rangle_R = \frac{1}{\sqrt{M}} \sum_{p=0}^{M-1} e^{-2\pi ipy/M} |p\rangle_R \end{equation}
and the state of the system becomes
\begin{equation}
	U_{QFT} U|\tilde{\bm{s}}\rangle_{12}|R\rangle_R = \frac{1}{M\sqrt{\bm{N}}} \sum_{\lambda} \sum_{y=0}^{M-1} \sum_{p=0}^{M-1} e^{2\pi i (l\lambda -p)y/M} |\lambda\rangle_1 | \lambda\rangle_2  |p\rangle_R
\end{equation}
Summing over $y$, we deduce
\begin{equation}
	U_{QFT} U|\tilde{\bm{s}}\rangle_{12}|R\rangle_R = \frac{1}{M\sqrt{\bm{N}}} \sum_{\lambda} \sum_{p=0}^{M-1} \frac{e^{2\pi i l\lambda } -1}{e^{2\pi i (l\lambda -p)/M} -1} |\lambda\rangle_1 | \lambda\rangle_2  |p\rangle_R
\end{equation}
The coefficients are strongly peaked at $p \approx l\lambda$. At the peaks, the coefficients are approximately equal to $M$. Therefore,
\begin{equation}
	U_{QFT} U|\tilde{\bm{s}}\rangle_{12}|R\rangle_R \approx \frac{1}{\sqrt{\bm{N}}} \sum_{\lambda}  |\lambda\rangle_1 | \lambda\rangle_2  |l\lambda\rangle_R
\end{equation}
A measurement of the register $R$ yields $p$ with probability
\begin{equation}
	\mathcal{P} (p) = \left\| {}_R\langle p|U_{QFT} U|\tilde{\bm{s}}\rangle_{12}|R\rangle_R \right\|^2 = \frac{1}{M^2\bm{N}} \sum_{\lambda} \left| \frac{e^{2\pi i l\lambda } -1}{e^{2\pi i (l\lambda -p)/M} -1}  \right|^2
\end{equation}
which completes the derivation of Eq.\ \eqref{eq:9}.

Approximately, the probability $\mathcal{P} (p)$ vanishes for all $p$, except at $p\approx l\lambda$. At the peaks, each eigenvalue contributes $1/\bm{N}$. Therefore, each peak is proportional to the multiplicity of the corresponding eigenvalue. In particular, for $p=l$, which corresponds to the eigenvalue of interest $\lambda = \xi = 1$, we have
\be \mathcal{P} (l) = \frac{\beta_k^{\eps,\eps'}}{\bm{N}} + \frac{1}{M^2 \bm{N}} \sum_{\lambda \ne 1} \frac{\sin^2 \pi l\lambda}{\sin^2 \frac{\pi l (\lambda -1)}{M}}
\ee
where $\beta_k^{\eps,\eps'}$ is the dimension of the kernel of the persistent Laplacian ($k$th persistent Betti number). In the limit $M\to\infty$, the sum over eigenvalues $\lambda \ne 1$ vanishes, and we obtain 
\be\beta_k^{\eps,\eps'} = \bm{N} \mathcal{P} (l)\ . \ee
\

\end{document}